# Three-dimensional crustal deformation analysis using physics-informed deep learning

Tomohisa Okazaki [a,*], Takeo Ito [b], Kazuro Hirahara [a,c], Ryoichiro Agata [d], Masayuki Kano [e], Naonori Ueda [a]

[a] RIKEN Center for Advanced Intelligence Project, 2-2 Hikaridai, Seika, Kyoto, 619-0288, Japan
[b] Nagoya University, Furo-cho, Chikusa-ku, Nagoya, 464-8602, Japan
[c] Kagawa University, Saiwai-cho 1-1, Takamatsu, Kagawa, 760-8521, Japan
[d] Japan Agency for Marine-Earth Science and Technology, 3173-25 Showa-machi, Kanazawa-ku, Yokohama, 236-0001, Japan
[e] Graduate School of Science, Tohoku University, 6-3 Aramaki-aza-aoba, Aoba-ku, Sendai, 980-8578, Japan

* Corresponding author at: RIKEN Center for Advanced Intelligence Project, 2-2 Hikaridai, Seika, Kyoto, 619-0288, Japan. *E-mail address:* tomohisa.okazaki@riken.jp (T. Okazaki).

**Abstract**

Earthquake-related phenomena such as seismic waves and crustal deformation impact broad regions, requiring large-scale modeling with careful treatment of artificial outer boundaries. Physics-informed neural networks (PINNs) have been applied to analyze wavefront propagation, acoustic and elastic waveform propagations, and crustal deformation in semi-infinite domains. In this study, we investigated the capability of PINNs for modeling earthquake crustal deformation in 3-D structures. To improve modeling accuracy, four neural networks were constructed to represent the displacement and stress fields in two subdomains divided by a fault surface and its extension. Forward simulations exhibited high accuracy for internal deformation but yielded errors for rigid motions, underscoring the inherent difficulty in constraining static deformation at an infinite distance. In the inversion analysis, fault slip distributions were estimated using surface observational data. Application to real data from the 2008 Iwate–Miyagi inland earthquake showed a fault slip consistent with previous studies, despite underestimation of the magnitude. This study demonstrates the capability of PINNs to analyze 3-D crustal deformation, thereby offering a flexible approach for large-scale earthquake modeling using real-world observations and crustal structures.

**1. Introduction**

Earthquakes, which are the rupture processes of source faults, generate transient seismic waves and permanent crustal deformation that can lead to natural hazards. Analyses of these phenomena involving forward simulations and inversions of the source processes from observational data have contributed to understanding earthquake processes and advancing disaster mitigation. In addition to traditional methods such as numerical solvers and Bayesian inversions, a deep learning approach of physics-informed neural networks (PINNs) has been gaining attention [1]. PINNs are a class of neural networks (NNs) designed for the forward and inverse modeling of partial differential equations (PDEs) and have been applied to diverse physical systems such as fluid dynamics [2,3] and solid mechanics [4,5] in scientific and engineering fields.

Earthquake-related phenomena such as seismic wave propagation and crustal deformation impact broad regions, typically ranging 10–1000 km, which necessitates their large-scale modeling. Because the Earth is even larger, a semi-infinite domain with artificial outer surfaces (OSs) is typically modeled, requiring careful treatment in numerical analysis. Such large-scale semi-infinite modeling is a salient feature of earthquake seismology. Another feature is that earthquake sources cannot be directly observed and must be inferred from the limited data collected on Earth's surface. Therefore, inversion analysis plays a crucial role by enabling the estimation of source processes or underground structures from observational data. The status of PINN applications in earthquake seismology is summarized in Table 1. Here, *n*-D and (*n*+1)-D represent time-independent and time-dependent problems of *n* spatial dimensions, respectively. Other studies have also modeled fault motions using rate and state dependent frictional law [6–8].

Wavefront propagation is one of the most widely studied research subjects, in which traveltime is calculated based on the eikonal equation. Since the pioneering work on local scales [9,10], its scope has expanded to 3-D regional and global scales [11,12]. Seismic tomography, which infers seismic velocity structures from observed traveltime data, was initiated from synthetic tests [13] and subsequently applied to real data [14]. From a computational perspective, the eikonal equation is the first-order PDE, and the first arrival time can be sequentially obtained from the source locations without explicitly considering the OSs.

Waveform propagation modeling is also actively conducted. Most studies have dealt with acoustic waves to facilitate analysis [15–17]. Full elastic waves were solved in [18], and synthetic inversions of acoustic waves were addressed in [16]. These studies analyzed (2+1)-D problems. Wave propagation has occasionally been solved in the frequency domain to reduce the input dimensions, such as acoustic waves in 2-D and 3-D structures [19–21], elastic waves in 2-D structures [22], and source imaging using 2-D real data [23]. Some studies suggested that no condition is required on OSs [16,20], whereas others introduced an absorbing boundary condition (BC) to ensure a unique solution [18,21].

Coseismic crustal deformation, that is, the static deformation of elastic media caused by earthquakes, has been analyzed under the assumption of 2-D antiplane [24,25] and 2-D inplane [26] strains. These analyses treated earthquake faults as finite surfaces, instead of point sources that are typically assumed in wavefront and waveform propagation modeling. This induced discontinuous internal boundaries, which increased the number of BCs. Additionally, vanishing BC was imposed on OSs to constrain the global behavior of solutions [26]. Consequently, the loss function of PINNs comprises many terms. Although the basic formulations of forward and inverse analyses have been established, these were limited to 2-D crustal structures and synthetic data. Therefore, this study aims to analyze the coseismic deformation in 3-D crustal structures using observational data and underground structure models, which are directly related to real-world applications. From a computational perspective, PINNs are applied to the second-order PDEs in 3-D semi-infinite domains with OSs.

The remainder of the article is organized as follows. In Section 2, we formulate the dislocation models for crustal deformation and PINN modeling. Section 3 presents forward analysis while Section 4 presents inversion analysis and considers actual data from the 2008 Iwate–Miyagi inland earthquake. Section 5 provides the implications of the results, before conclusions in Section 6.

**Table 1.** Applications of physics-informed neural networks to semi-infinite modeling in earthquake seismology

| Phenomenon | Wavefront propagation | Acoustic wave propagation | Elastic wave propagation | Crustal deformation |
| --- | --- | --- | --- | --- |
| Governing law | Eikonal equation | Wave equation | Elastic wave equation | Elastic equilibrium |
| Equation | First order PDE | Second order PDE | Second order PDE | Second order PDE |
| Spacetime | 3-D | T: (2+1)-D<br>F: 3-D | T: (2+1)-D<br>F: 2-D | 2-D → 3-D |
| Solution | Traveltime (scalar) | Pressure (scalar) | Displacement (vector) | Displacement (vector) |
| Earthquake | Point source | Point source | Point source | Finite fault |
| Inversion | Real data | T: Synthetic data<br>F: Real data | Not yet | Synthetic data → Real data |

Arrows (→) in the rightmost column indicate the contributions of this study. T, time domain; F, frequency domain. *n*-D and (*n*+1)-D represent time-independent and time-dependent problems of *n* spatial dimensions, respectively.

## 2. Methods

*2.1 Dislocation model*

Crustal deformation caused by earthquakes is mathematically described by a dislocation model, which considers defects (i.e., faults) in a continuum medium (i.e., the Earth's crust) and forms a system of PDEs. A dislocation model is specified by four components: surface topography defining the PDE domains, material properties represented by the PDE coefficients, fault geometry defining internal boundaries, and slip distribution represented by the displacement discontinuity (i.e., dislocation) on the internal boundaries. Forward analysis solves the deformation fields under specified conditions.

Coseismic crustal deformation obeys the equilibrium equation governing the spatial distribution of a symmetric stress tensor $\sigma_{ij}$:

$$\sigma_{ij,j} = 0. \tag{1}$$

Equation (1) is first-order PDEs, but the number of constraints (three) is less than the degrees of freedom (six). To fully constrain the physical system, stress must be related to displacement vector $u_i$ via a constitutive law. By assuming linear isotropic elastic media, the generalized Hooke's law yields:

$$\sigma_{ij} = \mu(u_{i,j} + u_{j,i}) + \lambda u_{k,k}\delta_{ij}, \tag{2}$$

where $\mu$ and $\lambda$ are elastic moduli (Lame's parameters) and $\delta_{ij}$ is Kronecker's delta. By substituting equation (2) into equation (1), we obtain a closed system of second-order PDEs.

In addition to the governing PDEs, BCs are defined for the four types of boundaries (Figure 1). Two of the boundaries, dislocation surface (DS) and free surface (FS), are physical. The DS is the fault surface that yields displacement discontinuity. The BCs comprise the displacement discontinuity and traction continuity:

$$u_i^+ - u_i^- = s_i, \tag{3}$$

$$\sigma_{ij}^+ n_j = \sigma_{ij}^- n_j, \tag{4}$$

where the superscripts $+$ and $-$ represent the opposite sides of the DS and $s_i$ and $n_i$ are the slip and normal vectors on the DS, respectively. The FS is the ground surface where the normal stress vanishes:

$$\sigma_{ij} n_j = 0. \tag{5}$$

The other two boundaries are artificial and arise from numerical modeling: contact surface (CS) and OS. Because NNs cannot directly represent discontinuous functions, the model domain is divided into two subdomains ($V_+$ and $V_-$) along the DS and the variables are represented by different NNs [26]. The CS is defined as the boundary between two subdomains, excluding the DS. Because it is not a physical boundary, displacement and stress must be continuous:

$$u_i^+ = u_i^-, \ \sigma_{ij}^+ = \sigma_{ij}^-. \tag{6}$$

Crustal deformation is mathematically formulated in semi-infinite domains, but numerical calculations are limited to finite model domains. Therefore, BCs must be imposed on the OS of the model domains to ensure proper behavior at an infinite distance. As earthquake deformation decays far from the fault, the displacement and stress should vanish on the OS:

$$u_i = 0, \ \sigma_{ij} = 0. \tag{7}$$

Note that the BCs (equations 3–7) are algebraic equations when displacement and stress are treated as independent variables.

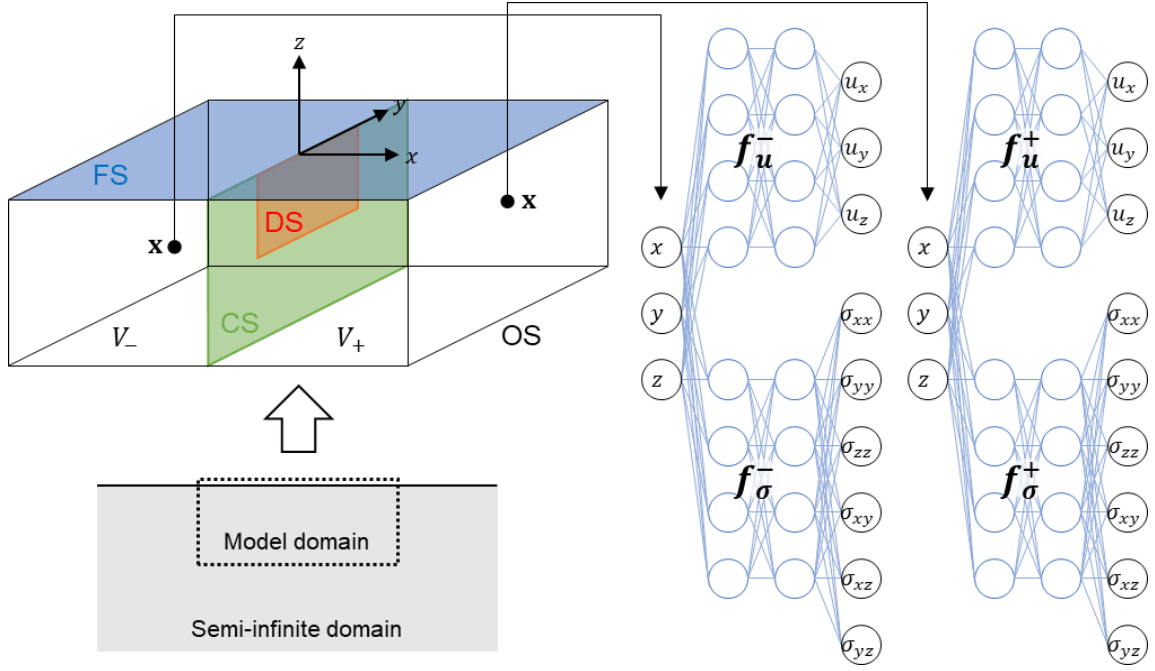

**Figure 1.** Schematic description of 3-D crustal deformation modeling using physics-informed neural networks. Left: A finite model domain defined in the semi-infinite domain. The model domain is bounded by the free surface (FS) and outer surface (OS) and divided into two subdomains ($V_+$ and $V_-$) along the dislocation surface (DS) and its extension called contact surface (CS). Right: Four neural networks corresponding to the two subdomains and two output variables, displacement $u_i$ and stress $\sigma_{ij}$.

*2.2 Physics-informed neural network (PINN)*

In the PINN framework, unknown solutions to PDEs are directly modeled by NNs with loss functions defined by PDEs and BCs. The input variables are the spatial coordinates $(x, y, z)$ in Cartesian coordinates, and output variables are the displacement $u_i$ and stress $\sigma_{ij}$. We construct four NNs ($f_u^+$, $f_\sigma^+$, $f_u^-$, and $f_\sigma^-$) corresponding to two subdomains and two output variables (Figure 1). Different NNs are used for $V_+$ and $V_-$ to represent the displacement discontinuity on the DS, which is a type of domain decomposition motivated by physical requirements rather than by computational efficiency [27,28]. Although a single NN for $u_i$ is sufficient in principle because $\sigma_{ij}$ is uniquely determined by equation (2), separate NNs are used for $u_i$ and $\sigma_{ij}$ to account for potentially significant difference in their magnitude and spatial attenuation properties [18].

The loss function is defined as the sum of the residuals of the PDEs and BCs (equations 1–7):

$$L^{(\text{for})} = L_{\text{pde}} + L_{\text{con}} + L_{\text{dsu}} + L_{\text{dst}} + L_{\text{fs}} + L_{\text{cs}} + L_{\text{os}}. \tag{8}$$

Here, each term is explicitly expressed as follows:

$$L_{\text{pde}} = (\sigma_{xx,x} + \sigma_{xy,y} + \sigma_{xz,z})^2 + (\sigma_{xy,x} + \sigma_{yy,y} + \sigma_{yz,z})^2 + (\sigma_{xz,x} + \sigma_{yz,y} + \sigma_{xz,z})^2, \tag{9}$$

$$\begin{aligned}L_{\text{con}} = &(\sigma_{xx} - 2\mu u_{x,x} - \lambda(u_{x,x} + u_{y,y} + u_{z,z}))^2 + (\sigma_{yy} - 2\mu u_{y,y} - \lambda(u_{x,x} + u_{y,y} + u_{z,z}))^2 \\ &+ (\sigma_{zz} - 2\mu u_{z,z} - \lambda(u_{x,x} + u_{y,y} + u_{z,z}))^2 + (\sigma_{xy} - \mu(u_{x,y} + u_{y,x}))^2 \\ &+ (\sigma_{xz} - \mu(u_{x,z} + u_{z,x}))^2 + (\sigma_{yz} - \mu(u_{y,z} + u_{z,y}))^2,\end{aligned} \tag{10}$$

$$L_{\text{dsu}} = (u_x^+ - u_x^- - s_x)^2 + (u_y^+ - u_y^- - s_y)^2 + (u_z^+ - u_z^- - s_z)^2, \tag{11}$$

$$\begin{aligned}L_{\text{dst}} = &((\sigma_{xx}^+ - \sigma_{xx}^-)n_x + (\sigma_{xy}^+ - \sigma_{xy}^-)n_y + (\sigma_{xz}^+ - \sigma_{xz}^-)n_z)^2 \\ &+ ((\sigma_{xy}^+ - \sigma_{xy}^-)n_x + (\sigma_{yy}^+ - \sigma_{yy}^-)n_y + (\sigma_{yz}^+ - \sigma_{yz}^-)n_z)^2 \\ &+ ((\sigma_{xz}^+ - \sigma_{xz}^-)n_x + (\sigma_{yz}^+ - \sigma_{yz}^-)n_y + (\sigma_{zz}^+ - \sigma_{zz}^-)n_z)^2,\end{aligned} \tag{12}$$

$$L_{\text{fs}} = (\sigma_{xx}n_x + \sigma_{xy}n_y + \sigma_{xz}n_z)^2 + (\sigma_{xy}n_x + \sigma_{yy}n_y + \sigma_{yz}n_z)^2 + (\sigma_{xz}n_x + \sigma_{yz}n_y + \sigma_{zz}n_z)^2, \quad (13)$$

$$L_{\text{cs}} = (u_x^+ - u_x^-)^2 + (u_y^+ - u_y^-)^2 + (u_z^+ - u_z^-)^2 + (\sigma_{xx}^+ - \sigma_{xx}^-)^2 + (\sigma_{xx}^+ - \sigma_{xx}^-)^2 + (\sigma_{xx}^+ - \sigma_{xx}^-)^2$$
$$+ (\sigma_{xx}^+ - \sigma_{xx}^-)^2 + (\sigma_{xx}^+ - \sigma_{xx}^-)^2 + (\sigma_{xx}^+ - \sigma_{xx}^-)^2, \quad (14)$$

$$L_{\text{os}} = u_x^2 + u_y^2 + u_z^2 + \sigma_{xx}^2 + \sigma_{yy}^2 + \sigma_{zz}^2 + \sigma_{xy}^2 + \sigma_{xz}^2 + \sigma_{yz}^2. \quad (15)$$

Here, the integral over domains and boundaries (or the sum over collocation points in practical implementation) are omitted for brevity of notation.

Until now, $u_i$ and $\sigma_{ij}$ are treated as independent variables. However, $\sigma_{ij}$ can be uniquely determined by $u_i$ through equation (2) in elastic media. Therefore, we can rewrite the equations above and construct the PINN model only using $u_i$. The two options for crustal deformation modeling are listed in Table 2. The displacement–stress (us) representation outputs $u_i$ and $\sigma_{ij}$ using the four NNs ($f_u^+$, $f_\sigma^+$, $f_u^-$, and $f_\sigma^-$), as shown in Figure 1. This has nine degrees of freedom and only approximately satisfies the constitutive law (equation 2). In contrast, the displacement (u) representation outputs $u_i$ using two NNs ($f_u^+$ and $f_u^-$), and $\sigma_{ij}$ in the loss function is calculated from equation (2). This only has three degrees of freedom and exactly satisfies the constitutive law. This apparently indicates that the 'u' representation is better. However, equation (2) contains derivatives that make the PDEs and some BCs higher-order differential equations. Higher-order derivatives considerably increase the computational cost of automatic differentiation [29] and complexity of the loss-function landscape [30], which can hamper efficient optimization. Previous studies on fluid dynamics and elastodynamics have suggested that lower-order representations are more efficient [2,18]. We therefore examine the practical efficiency and accuracy of these two representations in Sections 3.1 and 3.2.

**Table 2.** Two representations of crustal deformation modeling

| Representation | Displacement (u) | Displacement–stress (us) |
| --- | --- | --- |
| Variable (degrees of freedom) | $u_i$ (three) | $u_i$, $\sigma_{ij}$ (nine) |
| Governing equation | 2nd order PDE | 1st order PDE |
| Boundary condition | Algebraic, 1st order PDE | Algebraic equation |
| Constitutive law | Exact | Approximate |

*2.3 Inversion analysis*

Section 4 addresses slip inversion problems to estimate fault slip $s_i$ from displacement data ($x_i^{\text{data}}, u_i^{\text{data}}$) observed at discrete positions on the FS. In this problem setting, the displacement discontinuity (equation 3) is unknown and the corresponding loss term (equation 11) is removed from the loss function. Instead, we assume purely shear slips; that is, the displacement component normal to the DS is continuous:

$$u_i^+ n_i = u_i^- n_i. \quad (16)$$

Because observational data can constrain the global behavior of the solutions, the BC on the OSs can be omitted (see Section 3.2). Thus, the loss function is modified as follows:

$$L^{(\text{inv})} = L_{\text{pde}} + L_{\text{con}} + L_{\text{dsn}} + L_{\text{dst}} + L_{\text{fs}} + L_{\text{cs}} + L_{\text{data}}, \quad (17)$$

where the additional loss terms are defined by:

$$L_{\text{dsn}} = ((u_x^+ - u_x^-)n_x)^2 + ((u_y^+ - u_y^-)n_y)^2 + ((u_z^+ - u_z^-)n_z)^2, \quad (18)$$

$$L_{\text{data}} = (u_i - u_i^{\text{data}})^2. \quad (19)$$

The NN architecture is identical to that used in the forward analysis. Estimates of slip $s_i$ are obtained from the following relation:

$$s_i = u_i^+ - u_i^-, \tag{20}$$

where the right-hand side is given by the NN outputs.

*2.4 Implementation*

We used fully connected NNs with ten hidden layers, each containing 64 neurons, and the hyperbolic tangent (tanh) function as the activation function. The spatial coordinates were normalized using typical length scales before being input to the NNs. The OSs were set at a distance that was 100 times the fault depth in the forward analysis. The relative weights in the loss function were set to one for all terms. In every optimization step, collocation points were sampled densely near the fault but sparsely far from it to resolve the large deformation area while suppressing the computational cost. The batch sizes were 10,000 for the 3-D domains and 1000 for the 2-D boundaries. The Adam optimizer was used for training, where the initial learning rate was $10^{-3}$ and this was multiplied by 0.9 every 10,000 steps. The total number of steps was 500,000, except in Sections 3.1 and 3.2. All computations were performed using a single NVIDIA Tesla V100 GPU. The computational time typically ranged 60,000–70,000 s.

## 3. Forward Analysis

*3.1 Homogeneous half-space*

We consider a homogeneous half-space, for which semi-analytical solutions of coseismic deformation can be obtained [31]. We assume the Poisson material ($\mu = \lambda$) in elastic properties and consider planar faults of dip angles $\delta = 30°, 60°,$ and $90°$ (Figure 2a, b) expressed as:

$$x(y, z) = \tan(90° - \delta) \cdot z, \tag{21}$$

which defines the union of the DS and CS. Fault slips are imposed on this plane using bicubic spline functions:

$$s(y, z) = \phi(y)\phi(z), \tag{22}$$

$$\phi(y) = \begin{cases} 2(y+1)^3 & -1 \le y \le -0.5 \\ -6y^3 - 6y^2 + 1 & -0.5 \le y \le 0 \\ 6y^3 - 6y^2 + 1 & 0 \le y \le 0.5 \\ -2(y-1)^3 & 0.5 \le y \le 1 \end{cases}, \tag{23}$$

which have a maximum slip of 1 at the origin (Figure 2c). Because the fault slips occur in a bounded region ($-1 \le y \le 1, -1 \le z \le 0$), we define this region as the DS and its complement in the plane as the CS. Slip directions are set as reverse, normal, and left-lateral strike-slip motions for $\delta = 30°, 60°,$ and $90°$, respectively. Note that the physical variables are expressed as normalized nondimensional values.

    We compared the 'u' and 'us' representations, as shown in Table 2. The number of training steps for the 'u' representation was fixed at 250,000, whereas that for the 'us' representation was at 250,000 and 500,000 to compare the performances on the same training steps and approximately the same training time, respectively. To account for the variance owing to stochastic optimization, PINNs were trained on five random seeds for each model.

    The performances of these models are compared in Table 3. The 'u' representation is more accurate when comparing the same training steps, and the 'us' representation is competitive or more accurate when comparing the same training times. Therefore, the 'us' representation is slightly more efficient and accurate in this problem setting, although the difference is insignificant. The estimated surface displacement and stress for the reverse fault model are shown for the 'us' representation with 500,000 steps in Figure 3 and semi-analytical solutions in Figure 4. Note that, in half-space, $\sigma_{xz}, \sigma_{yz},$ and $\sigma_{zz}$ vanish on the FS owing to equation (5). These indicate that the PINNs are qualitatively accurate, including complex stress patterns. To compare the solutions

quantitatively, the surface displacements along the 1-D sections in Figure 2a are shown in Figure 5 (those for the normal and strike-slip faults are shown in Appendix A). The PINN solutions are accurate except for the $y = 1.5$ section, which is far from the DS and the deformation is small (approximately 1% of the maximum slip).

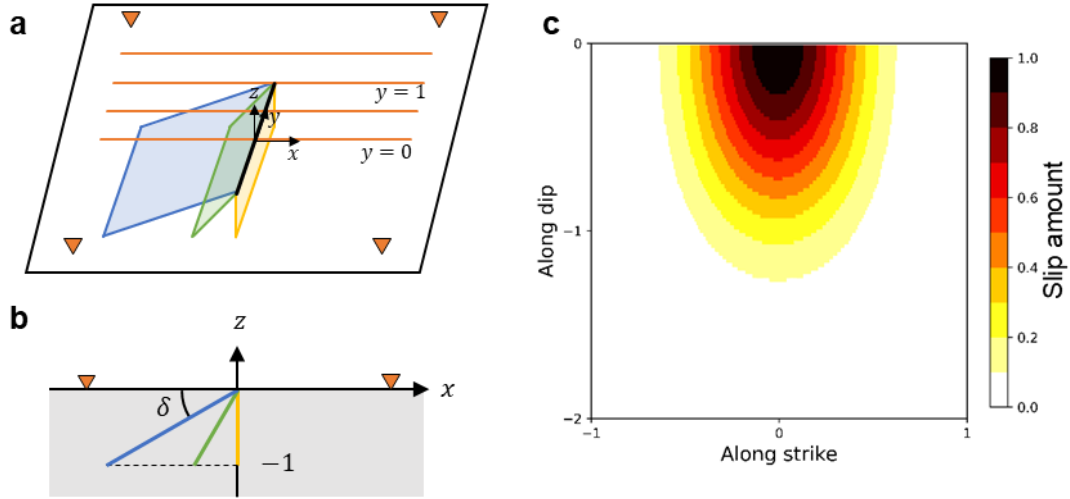

**Figure 2.** Assumed fault slip models. (a) Bird's eye and (b) cross-sectional views of fault geometry. The black parallelogram represents the ground surface. Blue, green, and yellow parallelograms represent subsurface planar faults with dip angles ($\delta$) of 30°, 60°, and 90°, respectively. The black line indicates the intersection of the ground and fault surfaces. Red triangles represent the positions of the supervised data in Section 3.2. Red lines indicate the cross sections in Figures 5 and 6. (c) Distribution of the slip amount on the fault with a dip angle of 30°. The top edge corresponds to the intersection with the ground surface.

**Table 3.** Estimation performance for the forward analysis

| Variable | Step | Time (s) | Relative $L_2$ error of surface displacements (%) | | |
| --- | --- | --- | --- | --- | --- |
| | | | Reverse | Normal | Strike-slip |
| u | 250,000 | 73759±3701 | **2.42±0.63** | **2.96±0.56** | 3.80±0.87 |
| us | 250,000 | 32243±699 | 4.24±1.15 | 4.68±1.24 | 5.05±1.96 |
| us | 500,000 | 64348±2301 | **2.27±0.75** | **2.67±0.70** | **2.62±0.85** |

Bold font indicates the smallest values within one standard deviation.

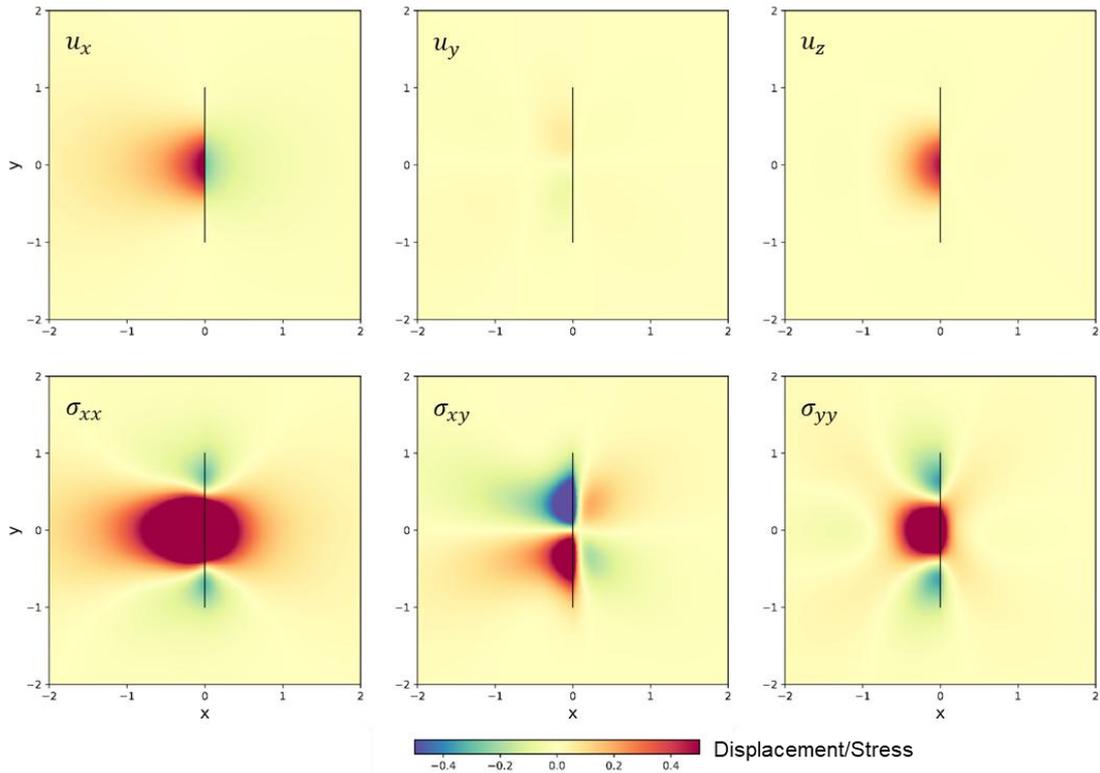

**Figure 3.** Estimated surface displacement (top) and stress (bottom) in the homogeneous half-space for the forward analysis. The results of the reverse faults are shown for the displacement–stress representation with 500,000 steps. Black lines represent the surface fault.

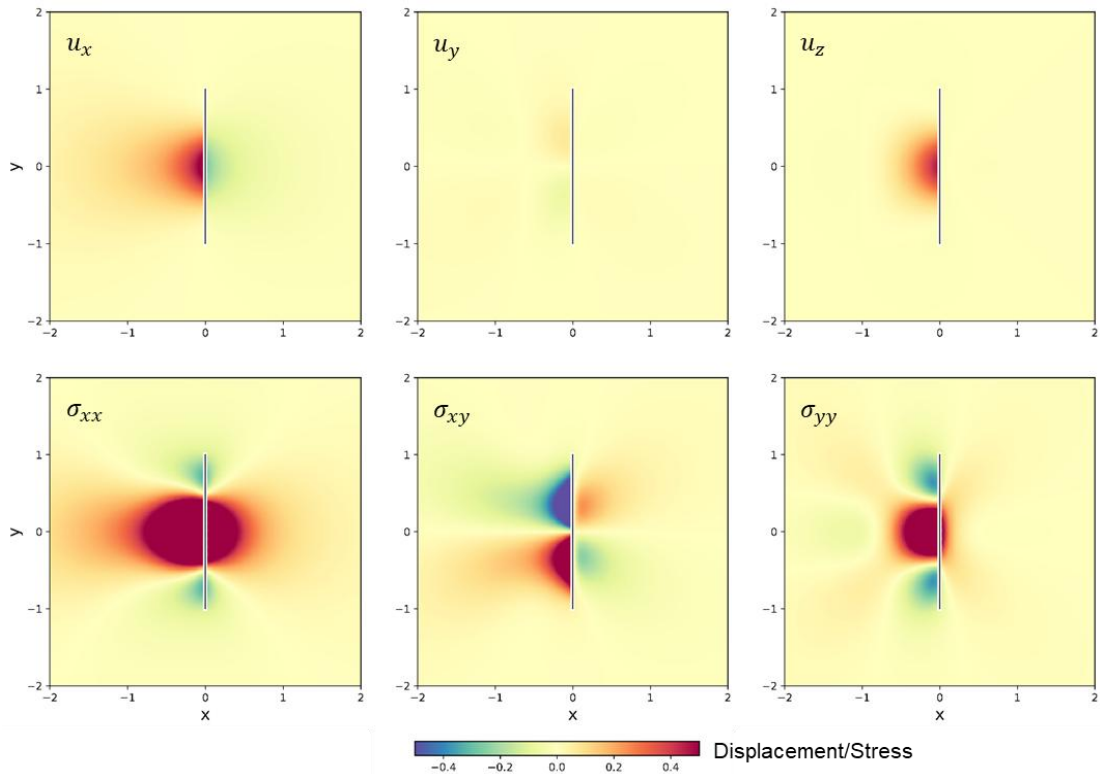

**Figure 4.** Semi-analytical surface displacement (top) and stress (bottom) in the homogeneous half-space. The solutions of the reverse faults are shown. Black lines represent the surface fault. Points near the fault showing numerical instability are not plotted.

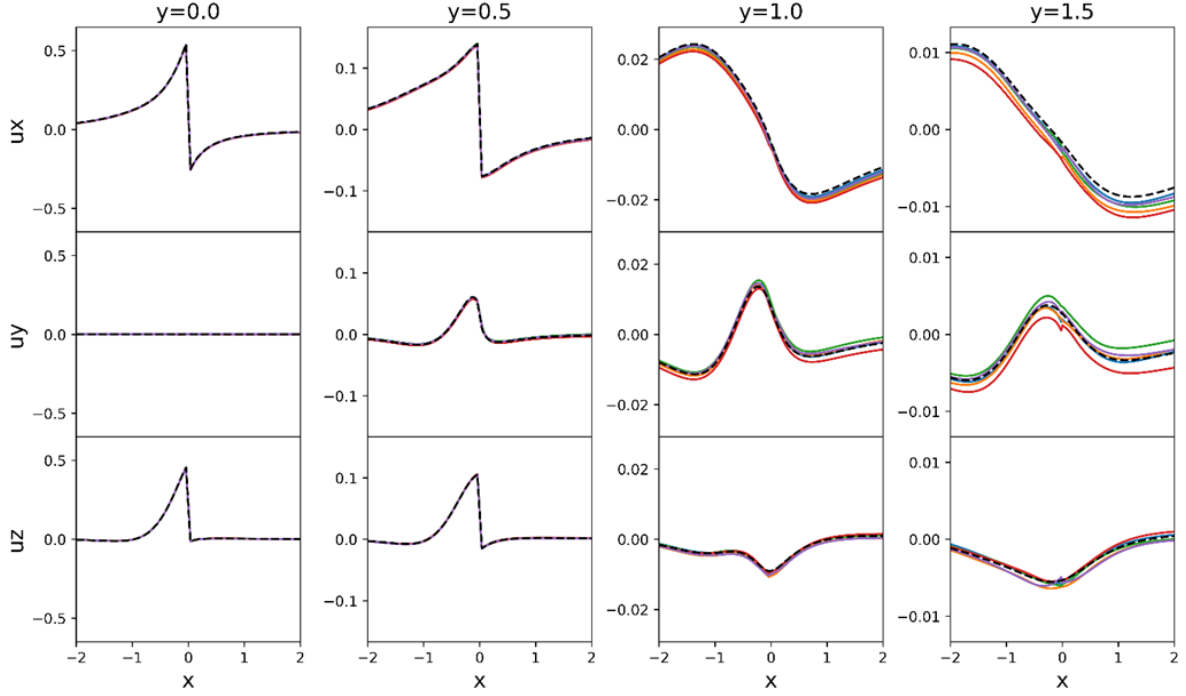

**Figure 5.** Cross sections of the estimated displacements in the homogeneous half-space for the forward analysis. The results of the reverse faults are shown for the displacement–stress representation with 500,000 steps. Colors represent different random seeds. Dashed lines represent the semi-analytical solutions. Note that the scales of the vertical axes are identical along columns but differ among columns.

*3.2 Difficulty in constraining rigid motions*

Although the results of the forward analysis are accurate near the fault ($y \leq 0.5$), significant errors are observed away from the fault ($y \geq 1.0$) (Figure 5). This discrepancy is essentially a constant shift, rather than a difference in shape. This is because equations (1)–(6) only constrain the internal deformation but leave the indeterminacy of rigid motion (i.e., translation and rotation). The rigid motion can be constrained only through the BC at infinity (equation 7); however, it is approximated at a finite distance in PINN modeling, inducing significant offsets in displacement.

To separate the errors resulting from the internal deformation and rigid motion, we constrain the rigid motion using the ground truth. Specifically, the semi-analytical solutions to displacement are given at four points $(x, y) = (\pm 2. \pm 2)$ (shown by red triangles in Figure 2a) as supervised data to fully constrain both the translational and rotational components (the corresponding loss term is denoted as $L_{\text{data}}$). In this case, the BC on the OS can be omitted, and the loss function is modified from equation 1 as follows:

$$L^{(\text{sup})} = L_{\text{pde}} + L_{\text{con}} + L_{\text{dsu}} + L_{\text{dst}} + L_{\text{fs}} + L_{\text{cs}} + L_{\text{data}}. \tag{24}$$

We refer to this problem setting as a supervised forward analysis, in which errors arise only from internal deformation. The PINNs were trained on five random seeds for each fault model.

The estimation results are presented in Table 4. The errors decrease significantly compared with those in the forward analysis, which suggests that the primary error sources arise from rigid motion. The reason for this is discussed in Section 5. The error reduction is higher in the 'us' representation, which achieves relative errors lower than 1%; The 1-D sections of the surface displacements in Figure 6 exhibit good agreement with the semi-analytical solutions (those for normal and strike-slip faults are shown in Appendix B). Therefore, we adopt the 'us' representation in subsequent sections.

**Table 4.** Estimation performance for the supervised forward analysis

| Variable | Step | Time (s) | Relative $L_2$ error of surface displacements (%) | | |
|---|---|---|---|---|---|
| | | | Reverse | Normal | Strike-slip |
| u | 250,000 | 74902±2511 | 1.17±0.40 | 1.67±0.32 | 1.66±0.56 |
| us | 250,000 | 33286±810 | 0.50±0.05 | 0.92±0.21 | 0.78±0.09 |
| us | 500,000 | 66711±2305 | **0.28±0.02** | **0.43±0.04** | **0.51±0.10** |

Bold font indicates the smallest values within one standard deviation.

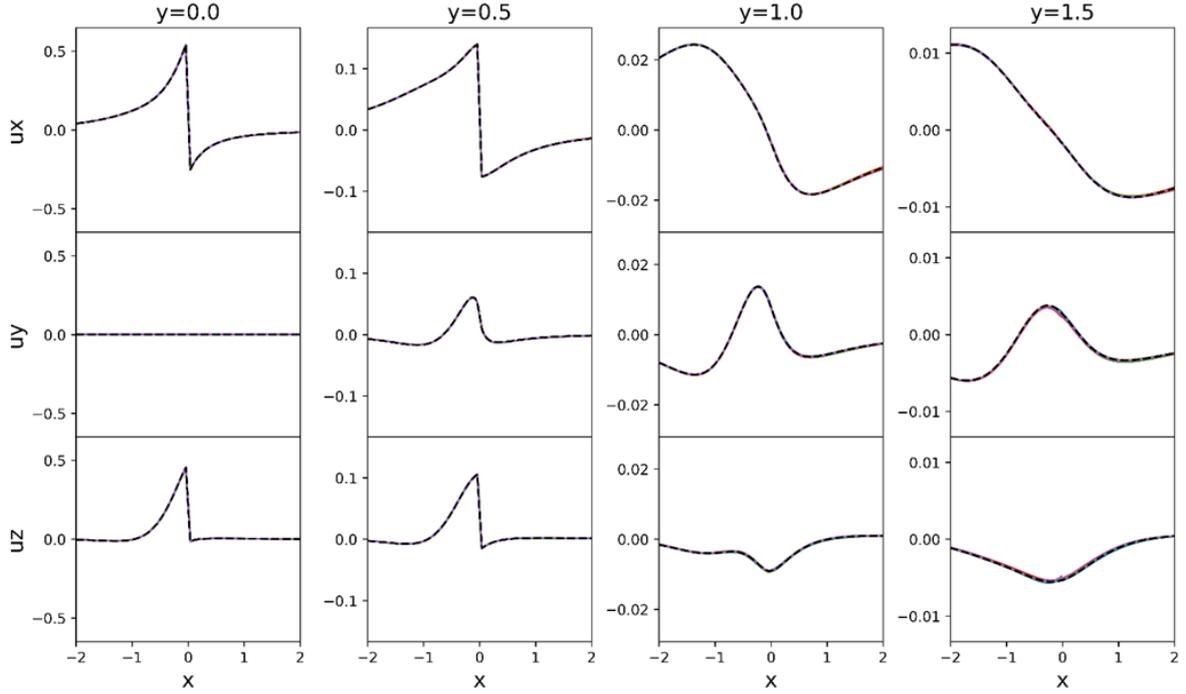

**Figure 6.** Cross sections of the estimated displacements in the homogeneous half-space for supervised forward analysis. The results of the reverse faults are shown for the displacement–stress representation with 500,000 steps. Colors represent different random seeds. Dashed lines represent the semi-analytical solutions. Note that the scales of the vertical axes are identical along columns but differ among columns.

*3.3 Heterogeneous structure*

PINNs can be adapted to heterogeneous structures with simple implementations, which are critical for wavefront and waveform simulations. Although crustal deformation is considered insensitive to small-scale heterogeneities because it can be regarded as a long-wavelength limit of seismic motions, large-scale topography and elastic heterogeneity can significantly influence the resultant deformation. To validate the performance of the PINN for heterogeneous structures, we consider the following large-scale structure. The surface topography is given by

$$z(x,y) = a \exp\left[-\frac{1}{2}(\mathbf{x}_2 - \mathbf{m}_2)^T \Sigma^{-1}(\mathbf{x}_2 - \mathbf{m}_2)\right], \tag{25}$$

$$\mathbf{x}_2 = \begin{pmatrix} x \\ y \end{pmatrix}, a = 0.5, \mathbf{m}_2 = \begin{pmatrix} -1 \\ 0 \end{pmatrix}, \Sigma = \begin{pmatrix} 1.5 & 0.5 \\ 0.5 & 1.5 \end{pmatrix}, \tag{26}$$

which represents a mountain-shaped terrain. The elastic modulus $\mu(=\lambda)$ is given by

$$\mu(x,y,z) = \mu_0 + \frac{\mu_1}{1+\exp\left(\frac{z-z_0}{z_h}\right)} + \mu_2 \exp\left[-\frac{1}{2l^2}(\mathbf{x} - \mathbf{m})^T(\mathbf{x} - \mathbf{m})\right], \tag{27}$$

$$\mu_0 = 1, \mu_1 = 2, z_0 = -2, z_h = 1, \ \mathbf{x} = \begin{pmatrix} x \\ y \\ z \end{pmatrix}, \mu_2 = -1, \ \mathbf{m} = \begin{pmatrix} -1 \\ 0 \\ -1 \end{pmatrix}, l = 2, \tag{28}$$

which increases with depth from $\mu_0$ to $\mu_0 + \mu_1$ and has a weak ball-shaped region at center $\mathbf{m}$. The ground structure is shown in Figure 7. We assume the same fault shape and slip distribution as the reverse fault in Section 3.1, although the intersection with the FS is different owing to topography.

The PINN estimations in the forward and supervised forward analyses are shown in Figure 8. Reference solutions were obtained using the finite element method with PyLith software (version 2.2.2) [32]. The accuracy is comparable to that of the homogeneous structure (Figures 5 and 6), which verifies the adaptability of PINNs to heterogeneous structures.

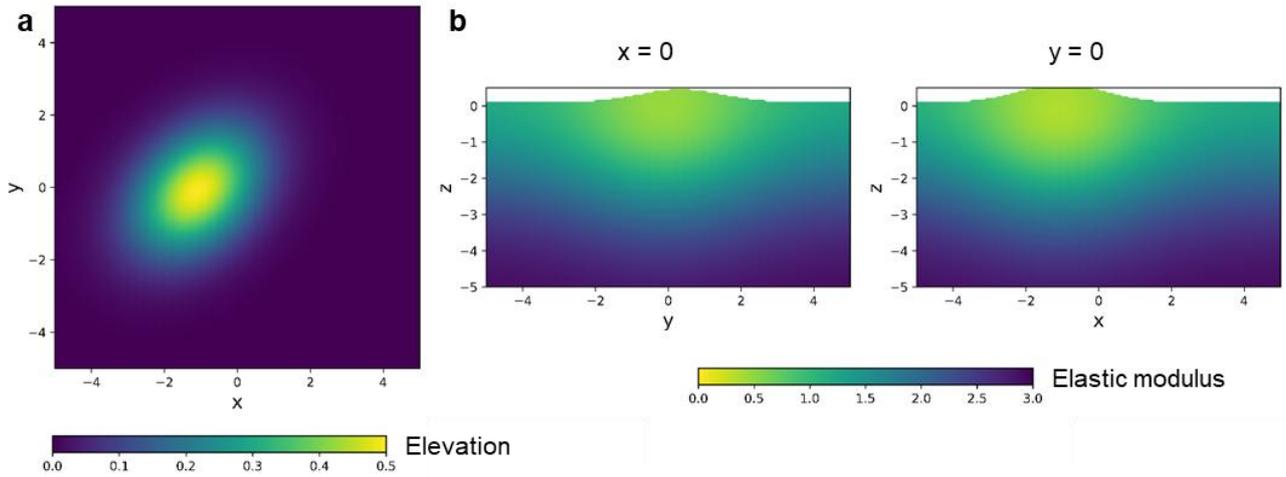

**Figure 7.** Heterogeneous crustal structure model. (a) Surface topography. (b) Elastic modulus along the $x = 0$ section (left) and the $y = 0$ section (right).

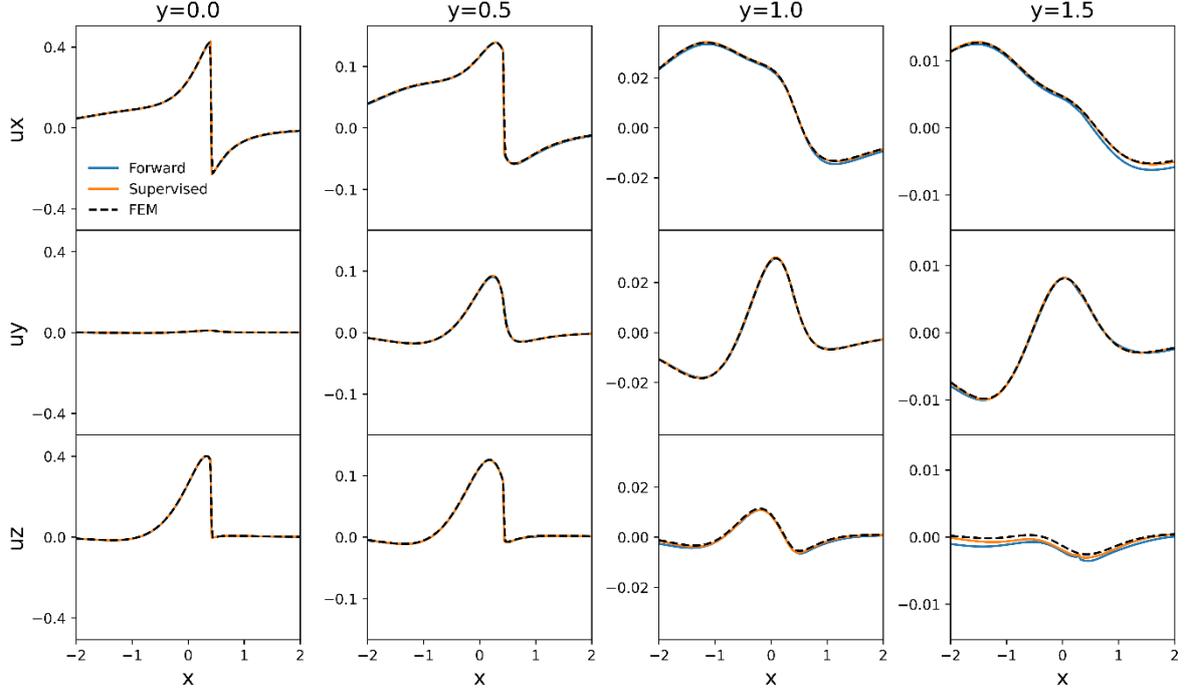

**Figure 8.** Cross sections of the estimated displacement in the heterogeneous structure. The results of the reverse faults are shown for the displacement–stress representation with 500,000 steps. Note that the scales of the vertical axes are identical along columns but differ among columns.

## 4. Inversion Analysis

### 4.1 Synthetic test

PINNs can address inversion analysis with minimal modifications to the loss function, as described in Section 2.3. We perform a synthetic test to evaluate the accuracy of the PINN-based inversions. We suppose the reverse fault in the homogeneous half-space in Section 3, but recover the dimensions by setting the unit values of the spatial coordinates and displacement as 10 km and 1 m, respectively, to introduce realistic noise levels. Synthetic observational data were produced as follows: $N = \{100, 64, 36, 16\}$ points were selected from the region $-20 \leq x, y \leq 20$ km on the FS and semi-analytical displacements were calculated, to which Gaussian noises of standard deviation $\sigma = \{0.5, 1, 2\}$ cm were added. In total, 12 datasets were prepared for inversion (Figure 9).

The accuracy of the inversions is evaluated using the relative $L_2$ errors of the fault slips $s_i$ (equation 20). To quantify the total scale of the earthquake, we use the seismic moment $M_0$, which is defined as

$$M_0 = \int_\Sigma \mu \|s_i\| d\Sigma \text{ (Nm)}. \tag{29}$$

Here, we assume $\mu = 30$ (GPa), following a convention. We also use the moment magnitude $M_w$, which is a logarithmic scale of $M_0$ defined by:

$$M_w = \frac{2}{3}(\log_{10} M_0(\text{Nm}) - 9.1). \tag{30}$$

The statistics and estimated slip distributions are presented in Table 5 and Figure 10, respectively. For dense and clean data, the relative $L_2$ errors of fault slips are <1%, demonstrating the high performance of the PINNs for slip inversions in ideal observational environments. When the data are sparser and noisier, the estimated slip distributions are distorted and the maximum slips are generally underestimated. In the worst case, the relative error exceeds 10% ($N = 36$ and $\sigma = 2$ cm). Notably, the maximum slip is underestimated by half because a large slip near the fault is not observed (Figure 9). However, the seismic moment is underestimated by only 8%, underscoring the stability of the PINNs in estimating the total slip amount.

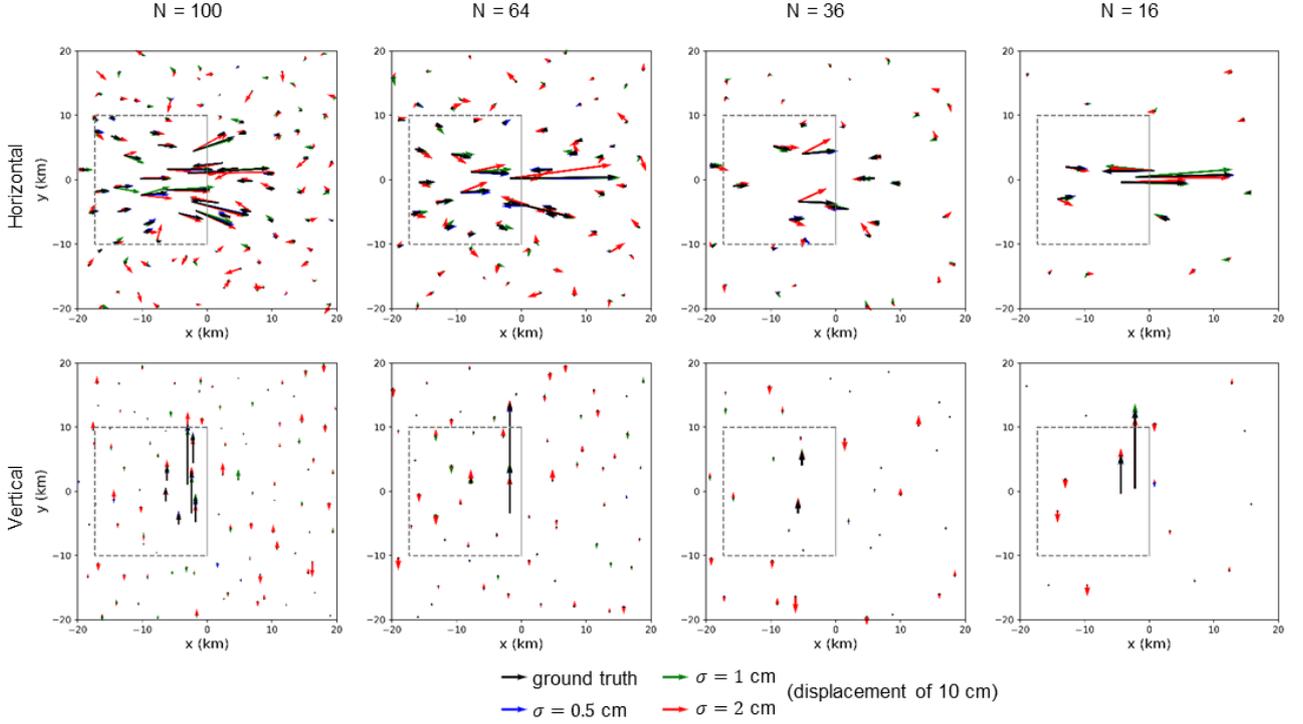

**Figure 9.** Synthetic displacement data. Colors correspond to added noise levels. Gray rectangles represent the surface projections of the fault (solid line indicates the intersection with the ground surface). $\sigma$, noise standard deviation; N, number of observations.

**Table 5.** Estimation performance for the synthetic inversion tests.

| $\sigma$ (cm) | N | Smax (m) | $M_0$ (Nm) | Mw | Error (%) |
|---|---|---|---|---|---|
| (Ground Truth) | | 1.00 | 1.66E+18 | 6.08 | (0) |
| 0.5 | 100 | 0.95 | 1.68E+18 | 6.08 | 0.41 |
| 0.5 | 64 | 1.00 | 1.64E+18 | 6.08 | 0.86 |
| 0.5 | 36 | 0.78 | 1.61E+18 | 6.07 | 3.28 |
| 0.5 | 16 | 0.94 | 1.82E+18 | 6.11 | 2.21 |
| 1 | 100 | 1.00 | 1.70E+18 | 6.09 | 0.70 |
| 1 | 64 | 0.96 | 1.84E+18 | 6.11 | 2.05 |
| 1 | 36 | 0.70 | 1.62E+18 | 6.07 | 5.56 |
| 1 | 16 | 0.99 | 1.79E+18 | 6.10 | 1.88 |
| 2 | 100 | 1.01 | 1.68E+18 | 6.08 | 2.28 |
| 2 | 64 | 0.92 | 1.86E+18 | 6.11 | 5.86 |
| 2 | 36 | 0.57 | 1.52E+18 | 6.05 | 13.84 |
| 2 | 16 | 0.85 | 2.02E+18 | 6.14 | 5.12 |

$\sigma$, noise standard deviation; N, number of observations; Smax, maximum slip; $M_0$, seismic moment; Mw, moment magnitude; Error, relative $L_2$ error of slip distribution.

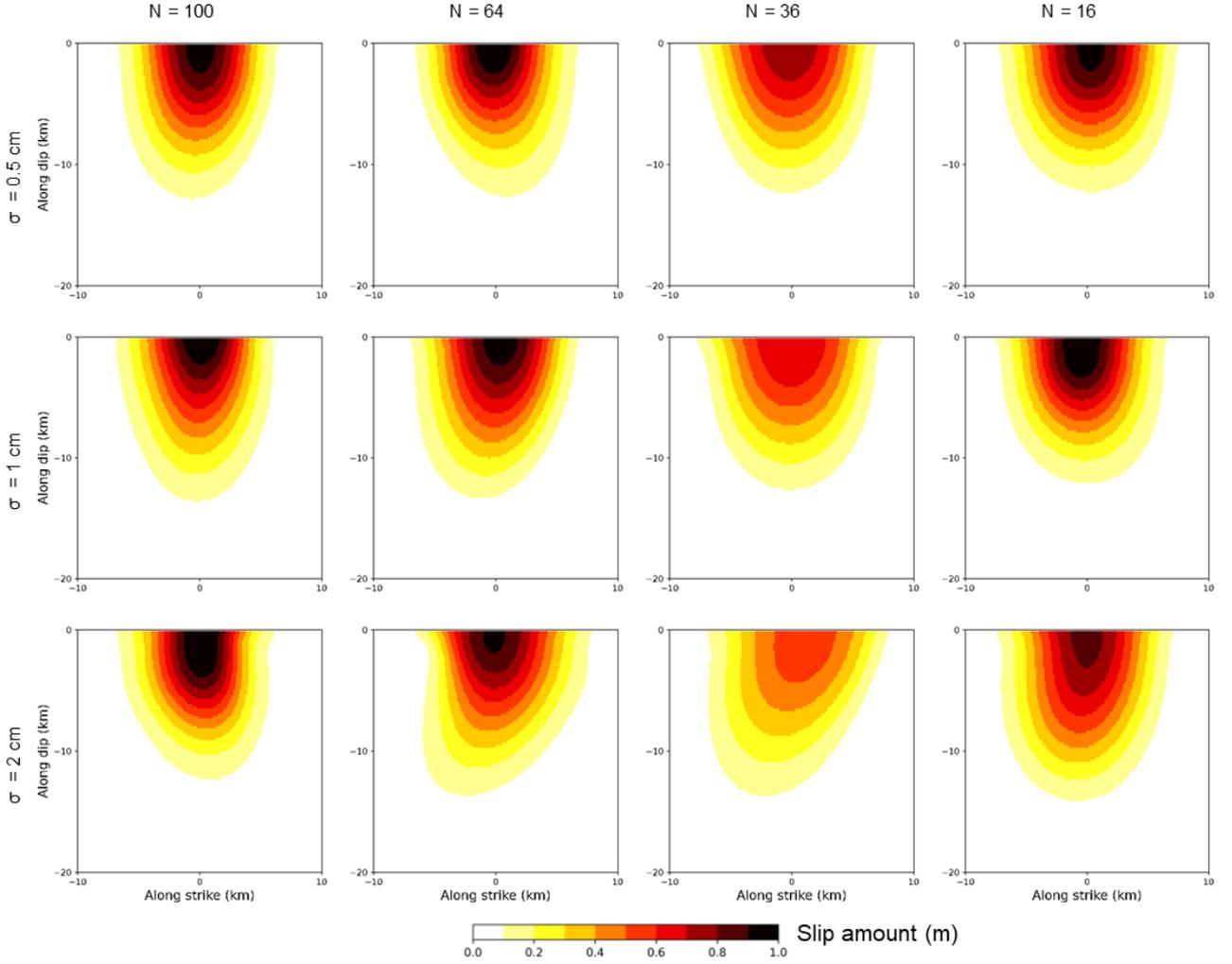

**Figure 10.** Estimated slip distributions in the synthetic test. The ground truth is shown in Figure 2c. $\sigma$, noise standard deviation; N, number of observations.

*4.2 Real data analysis of the 2008 Iwate–Miyagi inland earthquake*

We analyze the 2008 Iwate–Miyagi inland earthquake in northeast Japan known for the extreme ground motions recorded in the vicinity of the surface fault, which reached four times the gravitational acceleration [33]. We estimate the fault slip of this event from the actual observational data using PINNs.

Global Navigation Satellite System (GNSS) time-series data were obtained from the F5 solutions of GEONET. Coseismic displacements were calculated as the difference in coordinates between the average of ten days before the event and the day of the event. We set the fault plane based on [34], which estimated strike and dip angles of 198° and 31°, respectively, and the length and width of 20 and 9 km, respectively. We assumed the same fault location but extended the length and depth to 30 and 10 km (resulting in a width of 19.416 km), respectively, because a uniform slip was assumed in [34]. The model region was set as a 160 × 160 km square whose origin and axis coincided with the center of the surface fault (140.887°E, 38.961°N) and its strike direction (N18°E), respectively (Figure 11a), which led to 51 observational sites for inversion (Figure 11b). A large displacement in the surface projection of the fault corresponds to the station recording extreme ground motions. Inversion was performed for five random seeds, and the average was used as the estimation result. The value of the loss function was $6.73 \times 10^{-7}$.

The estimated slip distribution is shown in Figure 12a. The slip direction is mainly reverse with a small left lateral strike-slip component. The maximum slip is 4.50 m, consistent with 3.5–6.2 m reported in previous studies using GNSS data [34,35] and strong motion data [36]. The seismic moment is $1.58 \times 10^{19}$ Nm, corresponding to $M_w$ 6.73 ($\mu$ = 30 GPa is assumed), which is smaller than $M_w$ 6.9 obtained in previous studies [34–36]. The RMS residual from surface displacement data is $6.38 \times 10^{-5}$ m. This small value results from the soft constraints of physics: PINNs can fit discrete data by locally violating the PDEs near observation sites. These aspects will be further discussed in Section 5.

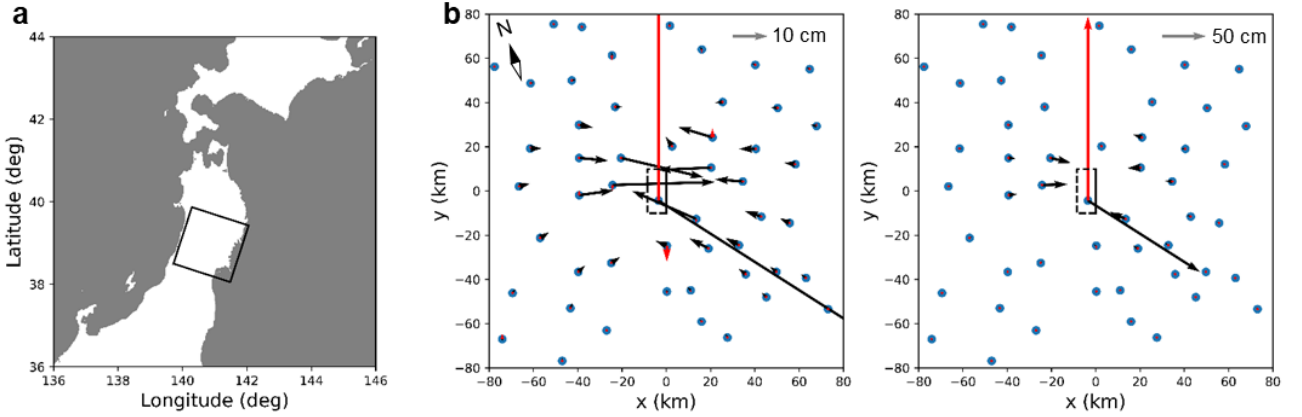

**Figure 11.** Observational data of the 2008 Iwate–Miyagi inland earthquake. (a) Geographical location of the target region in northeastern Japan. The rectangle indicates the model region in (b). (b) Observed coseismic displacements at two different displacement scales. The origin is located at (140.887°E, 38.961°N), and the $y$ axis is in the direction of N18°E. Black and red arrows represent horizontal and vertical displacements, respectively. Rectangles represent the surface projections of the fault (solid line indicates the intersection with the ground surface).

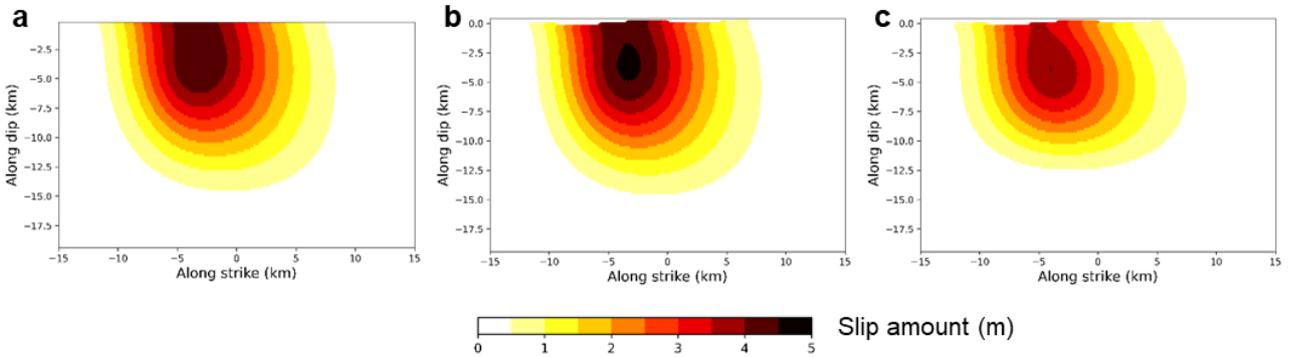

**Figure 12.** Estimated slip distributions of the 2008 Iwate–Miyagi inland earthquake in the (a) homogeneous half-space, (b) homogeneous structure with topography (Hom–Top) model, and (c) heterogeneous structure with topography (Het–Top) model.

*4.3 Realistic crustal structure*

In slip inversions, crustal structures are typically assumed to be homogeneous half-spaces, because heterogeneities are considered to have little influence on crustal deformation (Section 3.3). To examine their impact on slip estimation, we incorporate realistic topography and underground structure models into the inversion analysis. The Japan Integrated Velocity Structure Model (JIVSM) Version 1 [37] was used to represent the surface topography and elastic properties. This model assumes layered subsurface structures, with each layer having constant P-wave velocity $V_P$, S-wave velocity $V_S$, and density $\rho$. The heights of the upper surfaces of the layers are given as functions of the horizontal locations.

In this analysis, several shallow layers of JIVSM were combined to reduce the number of layers, and elastic moduli $\lambda$ and $\mu$ of each selected layer were calculated through the relation $V_P = \sqrt{(\lambda + 2\mu)/\rho}$ and $V_S = \sqrt{\mu/\rho}$ (Table 6). The surface topography $z_1 = f_{st}(x,y)$ and the subsurface layer boundary heights $(z_8, z_{15}, z_{16}) = f_{sl}(x,y)$ were fitted using NNs consisting with six hidden layers, each containing 16 neurons. The loss functions were mean-squared residuals from the JIVSM. A schematic description of the modeling and the obtained crustal structure are shown in Figure 13. The trained NNs can easily be incorporated into the subsequent PINN-based inversions: $f_{st}(x,y)$ was used to sample collocation points on the FS and to compute the normal vector $n_i$ appearing in $L_{fs}$ (equation 13) using automatic differentiation and $f_{sl}(x,y)$ was used to assign elastic moduli $\lambda$ and $\mu$ at given collocation points $x_i$ appearing in $L_{con}$ (equation 10). The inversions were performed in two crustal structure models: (i) a homogeneous structure with topography (Hom–Top) model using $f_{st}(x,y)$ but assuming constant elastic moduli; (ii) a heterogeneous structure with a topography (Het–Top) model using both $f_{st}(x,y)$ and $f_{sl}(x,y)$. The fault plane was the same as that in the half-space, although the intersection with the ground surface was different owing to topography. The values of the loss function were $7.98 \times 10^{-7}$ and $9.73 \times 10^{-5}$ for the Hom–Top and Het–Top models, respectively.

The estimated slip distributions are shown in Figure 12b, c. In the Hom–Top model, the maximum slip is 4.62 m and $M_w$ 6.73 (Figure 12b); $M_w$ is the same as, but the peak slip is larger than that in the half-space. In the flat structure, a large slip area is broadly distributed across the surface (Figure 12a). Because the center of the fault is located near mountain ridges (Figure 13b–d), incorporating the topography narrows the large slip area, making the peak sharper to produce the same total slip (Figure 12b). This underscores the importance of realistic crustal structures for inversions. Meanwhile, in the Het–Top model, the maximum slip is 4.00 m and $M_w$ 6.66 (Figure 12c), both smaller than those in the half-space. This is because the shallow portion deforms more easily and a smaller subsurface slip can produce the observed surface deformation [38]. However, the loss function remains large for the Het–Top model, suggesting difficulty in optimization owing to discontinuous changes in the elastic properties across the layer boundaries. Moreover, although a planar fault was assumed in this study, a nonplanar fault geometry has been suggested for this earthquake [39]. Reducing these modeling errors can lead to robust estimations of the source process.

**Table 6.** Selected subsurface layers and their physical parameters from the Japan Integrated Velocity Structure Model Version 1.

| Original layers | Selected layer | $V_P$ (m/s) | $V_S$ (m/s) | $\rho$ (g/cm³) | $\lambda$ (GPa) | $\mu$ (GPa) |
|---|---|---|---|---|---|---|
| 1–7 | 4 | 2.1 | 0.7 | 2.05 | 7.03 | 1.00 |
| 8–14 | 11 | 3.5 | 2 | 2.35 | 9.99 | 9.40 |
| 15 | 15 | 5.8 | 3.4 | 2.7 | 28.40 | 31.21 |
| 16 | 16 | 6.4 | 3.8 | 2.8 | 33.82 | 40.43 |

$V_P$, P-wave velocity; $V_S$, S-wave velocity; $\rho$, density; $\lambda$, $\mu$, Lame's parameters.

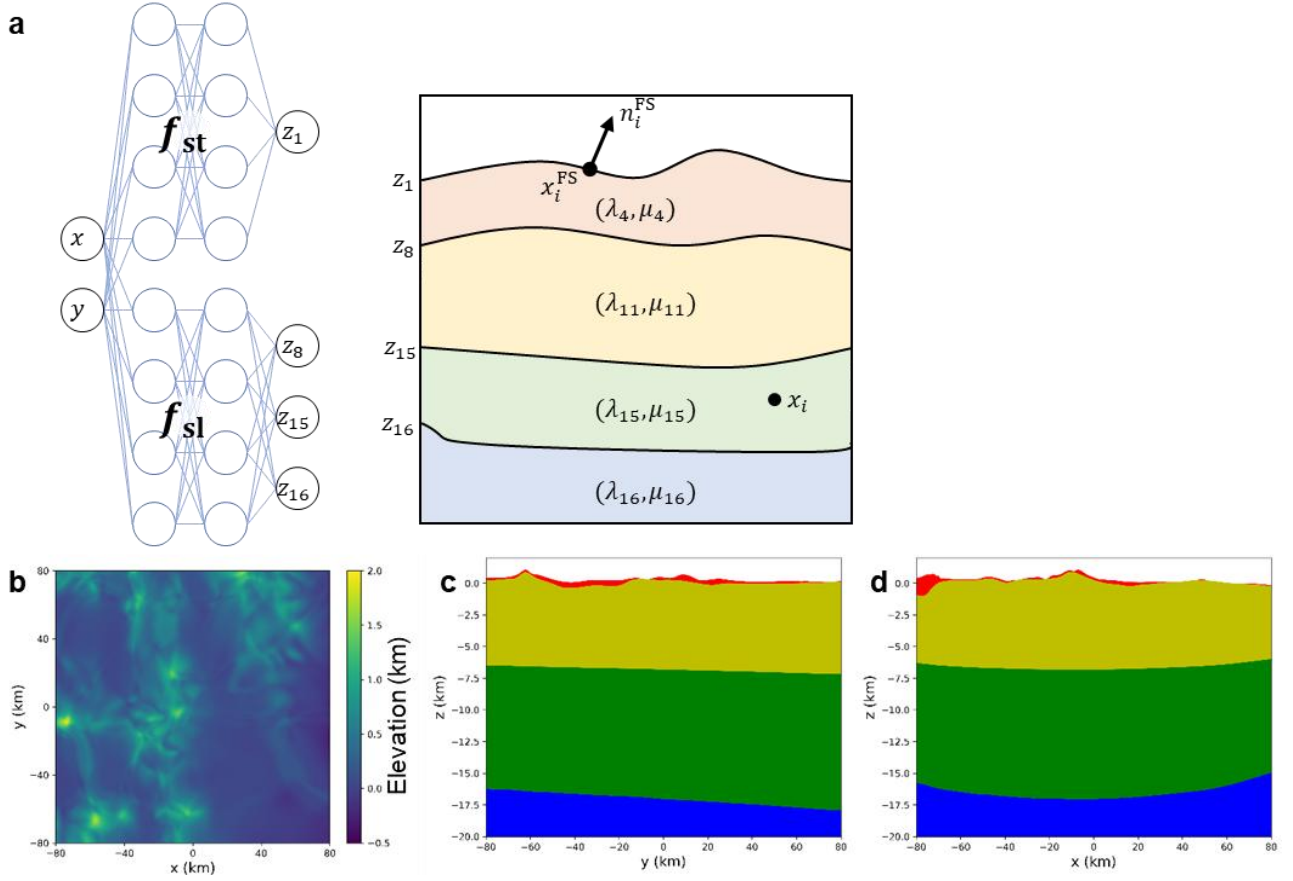

**Figure 13.** Crustal structure models for the 2008 Iwate–Miyagi inland earthquake. (a) Schematic description of the neural network (NN) modeling. Two NNs $z_1 = f_{st}(x, y)$ and $(z_8, z_{15}, z_{16}) = f_{sl}(x, y)$ are constructed to represent the surface topography and subsurface layer boundary heights, respectively. These are used in the subsequent inversion analysis. (b) Surface topography. (c) Underground structure along the $x = 0$ section. (d) Underground structure along the $y = 0$ section. In (c) and (d), colors (red, yellow, green, and blue) correspond to layers (Layers 4, 11, 15, and 16) in Table 6.

## 5. Discussion

In Section 3, the primary error sources are constraining the rigid motion, rather than resolving the internal deformation. This clarifies the characteristics of the BCs of coseismic deformation. At a finite distance, the displacement is not specified as the initial value or Dirichlet BC. However, its difference and derivatives (in the form of stress) are constrained as the displacement discontinuity (equation 3) and Neumann BCs (equations 4 and 5), respectively. The displacement is fixed only at an infinite distance (equation 7), from which the rigid motion components of the displacement near the fault must be determined. Generally, coseismic deformation obeys an elliptic PDE, for which BCs affect the entire domain and artificial OSs should be placed further away to suppress their influence. In contrast, seismic waves obey hyperbolic PDEs for which initial values are given and BCs affect locally owing to causality. This inherent indeterminacy suggests difficulties in forward simulations of crustal deformation, where true solutions are unavailable at any finite position. However, in the inversion analysis, the displacement is constrained by (occasionally noisy) observations at several positions, which determines rigid motion. Therefore, indeterminacy does not affect the performance of the data analysis.

In Section 4, although the PINNs fit noisy data almost perfectly, they did not produce unstable slip estimations without an explicit regularization term. The inherent bias of NNs converging to a regular solution during the optimization process is known as implicit regularization [40]. This property has also been observed in PINN applications where both data and physics are incomplete [41]. Some studies have suggested that solving the PDEs, instead of simply fitting the data, is crucial for effective regularization [8,26]. However, the estimated slip distribution of real data analysis (Figure 12) was much smoother than that obtained in a previous study using strong motion data [36]. This could be attributed to two reasons: limited resolution owing to the difference in analyzed data or insufficient optimization of the PINNs. These possibilities should be explored by accumulating the inversion results of other earthquake source processes using PINNs.

In Section 4.3, two NNs were trained on a heterogeneous crustal structure model (JIVSM) and incorporated into the PINN modeling (Figure 13). The use of NNs instead of tables of numerical values on grid points has several advantages. First, physical variables can be evaluated at arbitrary points in continuous domains, which is suitable for the random sampling of collocation points. Furthermore, the derivatives can be computed using automatic differentiation. Therefore, modeling using a common method can offer an efficient and streamlined analysis. For example, Agata et al. [42] constructed a subsurface velocity model using PINN-based seismic tomography and incorporated it into a hypocenter estimation, including the propagated uncertainty quantified by Bayesian inference [43]. Deep learning may provide a unified framework for modeling earthquake processes by constructing NNs of individual components such as background structures (e.g., material properties and earthquake faults) and deformation phenomena (e.g., fault motions, seismic waves, crustal deformation, and tsunamis). This strategy could be also applied to other scientific and engineering problems.

## 6. Conclusions

This study developed PINNs for crustal deformation caused by earthquakes in 3-D semi-infinite domains, which presents a computational challenge in geophysical problems. Four NNs were used to represent the displacement and stress fields in the two subdomains divided by fault surfaces because modeling the displacement and stress fields separately exhibited higher accuracy than modeling displacement fields only. In the forward analysis, the PINNs successfully solved the internal deformation, indicating their representation and optimization ability in 3-D semi-infinite modeling. However, major errors resulted from rigid-motion components, which suggests difficulty in constraining the BCs at an infinite distance. Although this poses a challenge in forward simulations, it can be remedied in an inversion analysis because the observational data can constrain rigid motions. In the inversion analysis, fault slip distributions were estimated from surface displacement data using PINNs. Synthetic tests indicated that the total slip amounts were stably estimated, although the peak slip was underestimated for sparse and noisy observations. Subsequently, the PINNs were applied to actual earthquake deformation data for the first time. The estimated slip distributions of the 2008 Iwate–Miyagi inland earthquake were generally consistent with the results of previous studies, despite the underestimation of the magnitude. A realistic crustal structure model was also incorporated into the inversion. These results demonstrates the capability of PINNs to analyze 3-D crustal deformation, which can enable large-scale earthquake modeling using real-world observations and crustal structures.

**CRediT authorship contribution statement**

**Tomohisa Okazaki:** Conceptualization, Formal analysis, Funding acquisition, Methodology, Software, Visualization, Writing—original draft. **Takeo Ito:** Data curation, Formal analysis, Supervision, Writing—review and editing. **Kazuro Hirahara:** Formal analysis, Supervision, Writing—review and editing. **Ryoichiro Agata:** Supervision, Writing—review and editing. **Masayuki Kano:** Supervision, Writing—review and editing. **Naonori Ueda:** Project administration, Supervision, Writing—review and editing.

**Acknowledgments**

This study was supported by the Grant-in-Aid for Scientific Research (B) (Kakenhi No. 25K01084) and the Grant-in-Aid for Scientific Research (C) (Kakenhi No. 23K03552) from the Ministry of Education, Culture, Sports, Science and Technology (MEXT) of Japan. This study was also supported by MEXT, under its The Third Earthquake and Volcano Hazards Observation and Research Program (Earthquake and Volcano Hazard Reduction Research).

**Declaration of competing interest**

The authors declare that they have no known competing interests.

**Data availability**

The GNSS data of GEONET are publicly available from the website of the Geospatial Information Authority of Japan (https://terras.gsi.go.jp/).

**Appendix A. Additional figures on forward analysis**

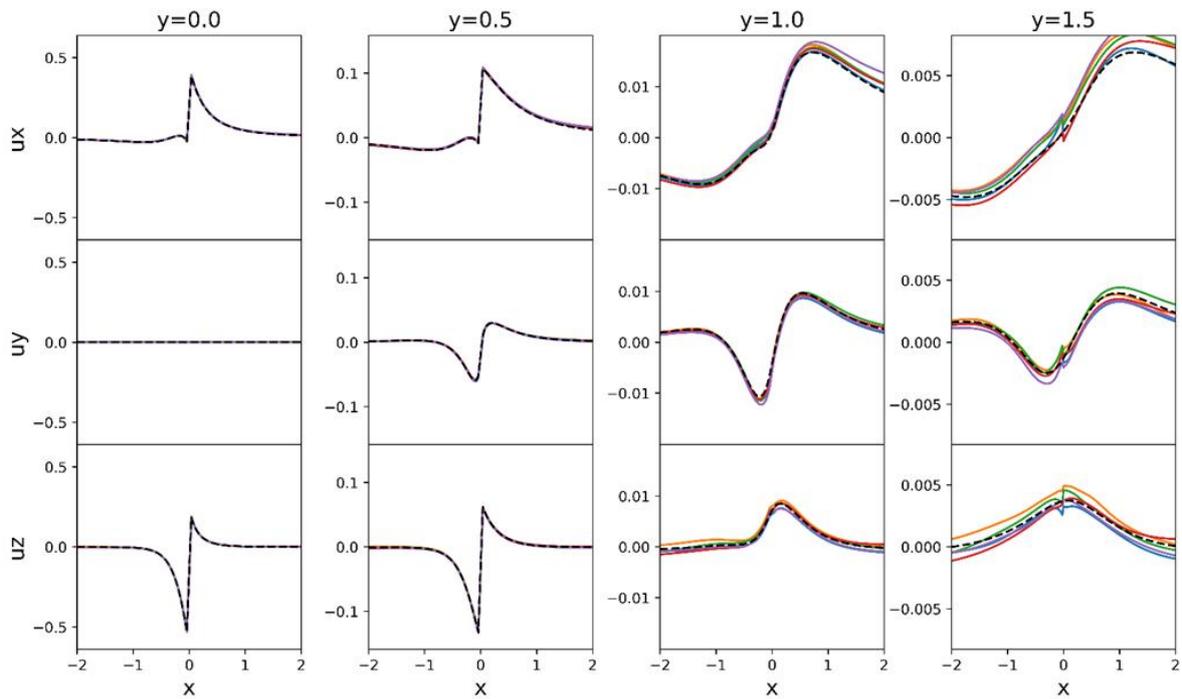

**Figure A1.** Cross sections of the estimated displacements in the homogeneous half-space for the forward analysis. The results of the normal faults are shown for the displacement–stress representation with 500,000 steps. Colors represent different random seeds. Dashed lines represent the semi-analytical solutions. Note that the scales of the vertical axes are identical along columns but differ among columns.

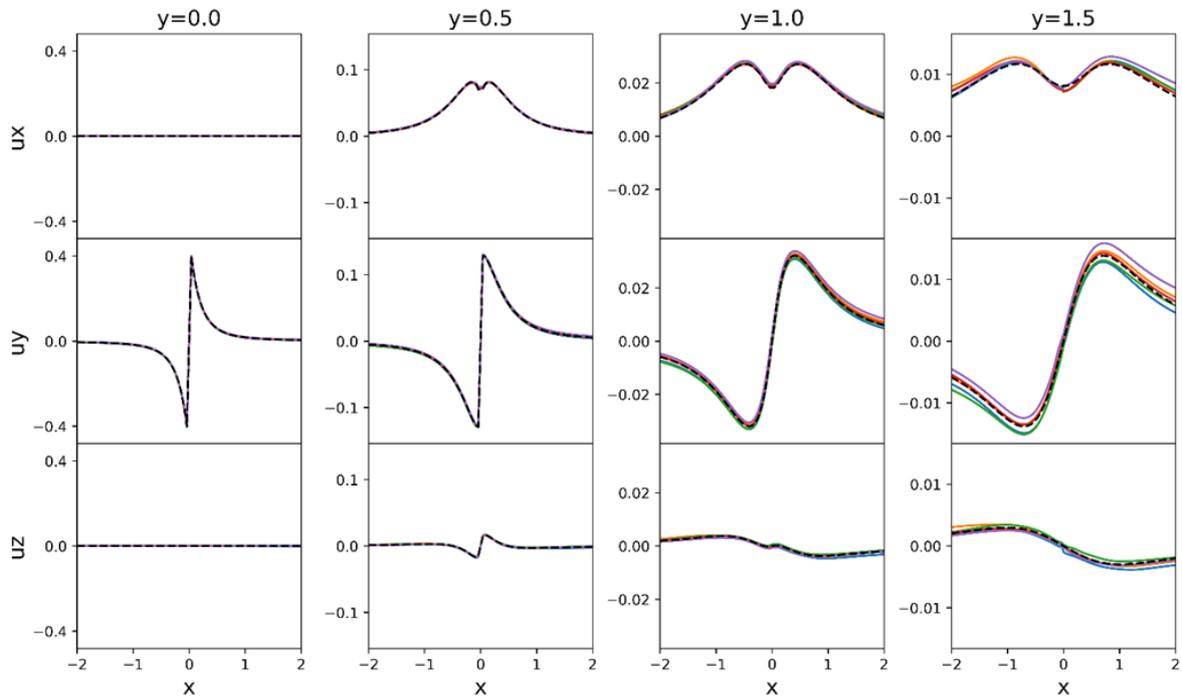

**Figure A2.** The same as Figure A1 but for the strike-slip faults.

**Appendix B. Additional figures on supervised forward analysis**

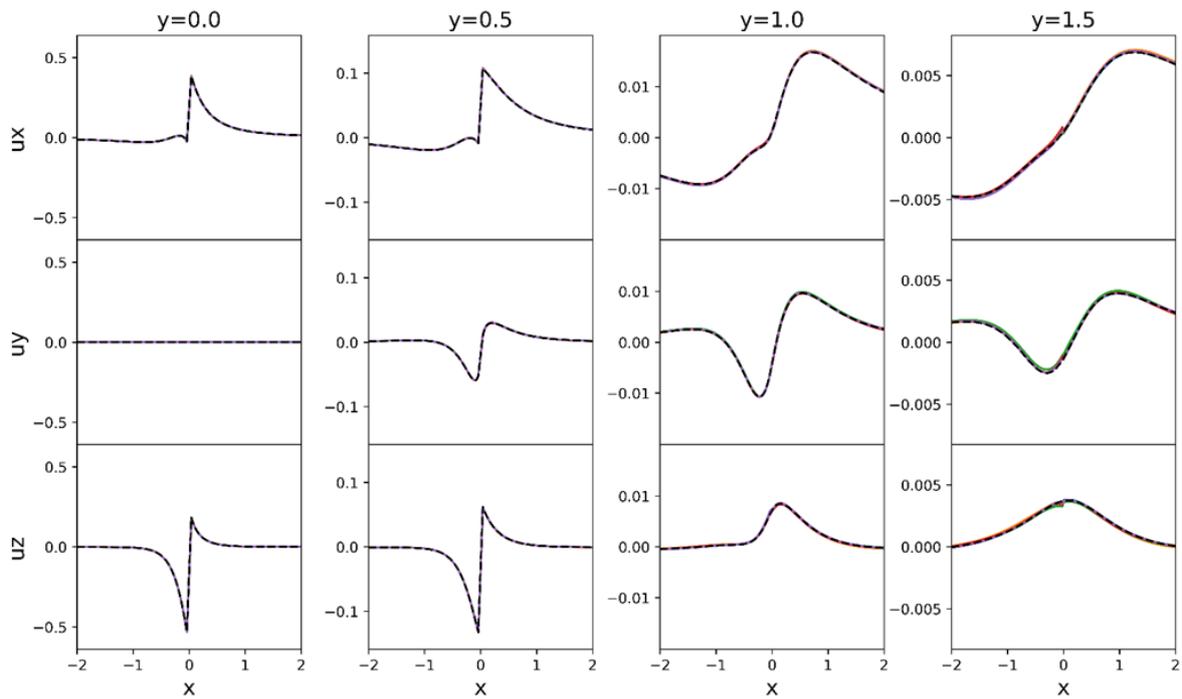

**Figure B1.** Cross sections of the estimated displacements in the homogeneous half-space for the supervised forward analysis. The results of the normal faults are shown for the displacement–stress representation with 500,000 steps. Colors represent different random seeds. Dashed lines represent the semi-analytical solutions. Note that the scales of the vertical axes are identical along columns but differ among columns.

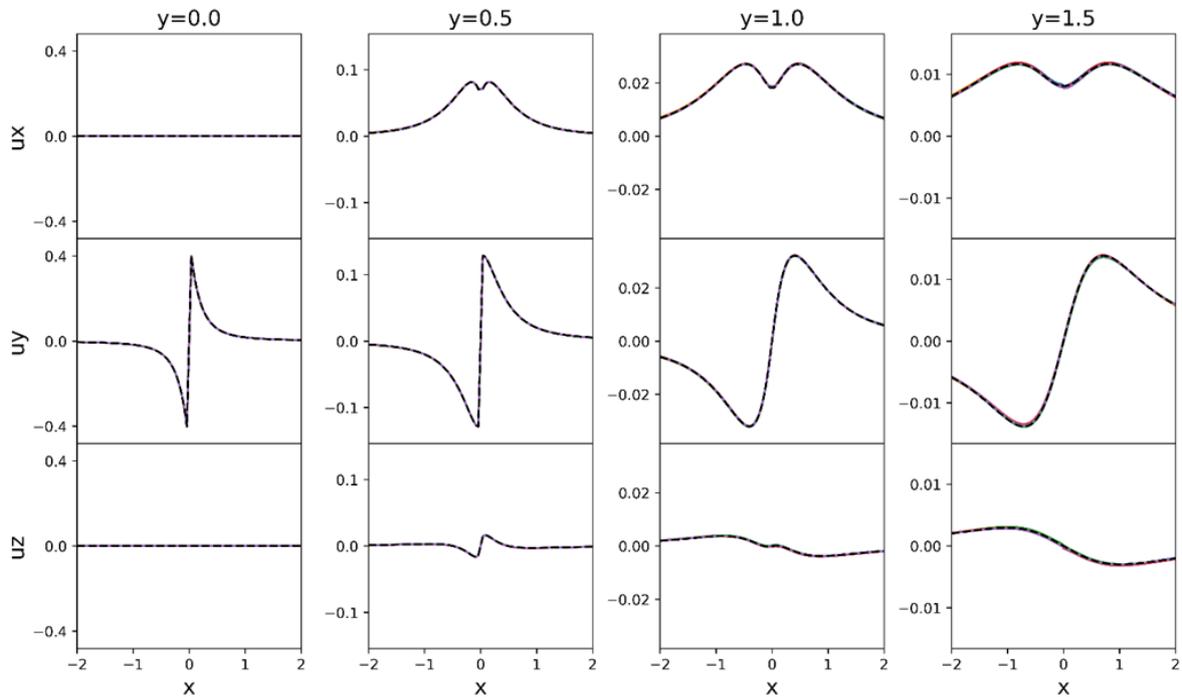

**Figure B2.** The same as Figure B1 but for the strike-slip faults.

**References**

[1] Raissi, M., Perdikaris, P., & Karniadakis, G. E. (2019). Physics-informed neural networks: A deep learning framework for solving forward and inverse problems involving nonlinear partial differential equations. *Journal of Computational Physics*, 378, 686–707. https://doi.org/10.1016/j.jcp.2018.10.045

[2] Rao, C., Sun, H., & Liu, Y. (2020). Physics-informed deep learning for incompressible laminar flows. *Theoretical and Applied Mechanics Letters*, 10(3), 207-212. https://doi.org/10.1016/j.taml.2020.01.039

[3] Sun, L., Gao, H., Pan, S., & Wang, J. X. (2020). Surrogate modeling for fluid flows based on physics-constrained deep learning without simulation data. *Computer Methods in Applied Mechanics and Engineering, 361*, 112732. https://doi.org/10.1016/j.cma.2019.112732

[4] Abueidda, D. W., Lu, Q., & Koric, S. (2021). Meshless physics-informed deep learning method for three-dimensional solid mechanics. *International Journal for Numerical Methods in Engineering*, 122(23), 7182–7201. https://doi.org/10.1002/nme.6828

[5] Haghighat, E., Raissi, M., Moure, A., Gomez, H., & Juanes, R. (2021). A physics-informed deep learning framework for inversion and surrogate modeling in solid mechanics. *Computer Methods in Applied Mechanics and Engineering*, 379, 113741. https://doi.org/10.1016/j.cma.2021.113741

[6] Fukushima, R., Kano, M, & Hirahara, K. (2023). Physics-informed neural networks for fault slip monitoring: simulation, frictional parameter estimation, and prediction on slow slip events in a spring-slider system. *Journal of Geophysical Research: Solid Earth*, 128(12), e2023JB027384. https://doi.org/10.1029/2023JB027384

[7] Rucker, C. and Erickson, B. A. (2024). Physics-informed deep learning of rate-and-state fault friction. *Computer Methods in Applied Mechanics and Engineering*, 430, 117211. https://doi.org/10.1016/j.cma.2024.117211

[8] Fukushima, R., Kano, M., Hirahara, K., Ohtani, M., Im, K., & Avouac, J.-P. (2025). Physics-informed deep learning for estimating the spatial distribution of frictional parameters in slow slip regions. *Journal of Geophysical Research: Solid Earth*, 130(5), e2024JB030256. https://doi.org/10.1029/2024JB030256

[9] Smith, J. D., Azizzadenesheli, K., & Ross, Z. E. (2021). EikoNet: Solving the eikonal equation with deep neural networks. *IEEE Transactions on Geoscience and Remote Sensing*, 59(12), 10685–10696. https://doi.org/10.1109/TGRS.2020.3039165

[10] Waheed, U. B., Haghighat, E., Alkhalifah, T., Song, C., & Hao, Q. (2021). PINNeik: Eikonal solution using physics-informed neural networks. *Computers & Geosciences*, 155, 104833. https://doi.org/10.1016/j.cageo.2021.104833

[11] Smith, J. D., Ross, Z. E., Azizzadenesheli, K. and Muir, J. B. (2022). HypoSVI: hypocentre inversion with Stein variational inference and physics informed neural networks, *Geophysical Journal International*, 228(1), 698–710. https://doi.org/10.1093/gji/ggab309

[12] Taufik, M. H., Waheed, U. B. and Alkhalifah, T. A. (2023). A neural network based global traveltime function (GlobeNN), *Scientific Reports*, 13(1), 7179. https://doi.org/10.1038/s41598-023-33203-1.

[13] Waheed, U. B., Alkhalifah, T., Haghighat, E., Song, C., & Virieux, J. (2021). PINNtomo: Seismic tomography using physics-informed neural networks. Preprint at https://doi.org/10.48550/arXiv.2104.01588

[14] Chen, Y., de Ridder, S. A., Rost, S., Guo, Z., Wu, X., & Chen, Y. (2022). Eikonal tomography with physics-informed neural networks: Rayleigh wave phase velocity in the northeastern margin of the Tibetan Plateau. *Geophysical Research Letters*, 49(21), e2022GL099053. https://doi.org/10.1029/2022GL099053

[15] Moseley, B., Markham, A. & Nissen-Meyer, T. (2020). Solving the wave equation with physics-informed deep learning. Preprint at https://doi.org/10.48550/arXiv.2006.11894

[16] Rasht-Behesht, M., Huber, C., Shukla, K., & Karniadakis, G. E. (2022). Physics-informed neural networks (PINNs) for wave propagation and full waveform inversions. *Journal of Geophysical Research: Solid Earth*, 127, e2021JB023120. https://doi.org/10.1029/2021JB023120

[17] Ding, Y., Chen, S., Li, X., Wang, S., Luan, S., & Sun, H. (2023). Self-adaptive physics-driven deep learning for seismic wave modeling in complex topography. *Engineering Applications of Artificial Intelligence*, 123, 106425. https://doi.org/10.1016/j.engappai.2023.106425

[18] Ren, P., Rao, C., Chen, S., Wang, J. X., Sun, H., & Liu, Y. (2024). SeismicNet: Physics-informed neural networks for seismic wave modeling in semi-infinite domain. *Computer Physics Communications*, 295, 109010. https://doi.org/10.1016/j.cpc.2023.109010

[19] Alkhalifah, T., Song, C., bin Waheed, U., & Hao, Q. (2021). Wavefield solutions from machine learned functions constrained by the Helmholtz equation. *Artificial Intelligence in Geosciences*, 2, 11–19. https://doi.org/10.1016/j.aiig.2021.08.002

[20] Song, C., Alkhalifah, T., & Waheed, U. B. (2021). Solving the frequency-domain acoustic VTI wave equation using physics-informed neural networks. *Geophysical Journal International*, 225(2), 846–859. https://doi.org/10.1093/gji/ggab010

[21] Chai, X., Gu, Z., Long, H., Liu, S., Cao, W., & Sun, X. (2024). Practical Aspects of Physics-informed neural networks applied to solve frequency-domain acoustic wave forward problem. *Seismological Research Letters*, 95(3), 1646–1662. https://doi.org/10.1785/0220230297

[22] Song, C., Liu, Y., Zhao, P., Zhao, T., Zou, J., & Liu, C. (2023). Simulating multicomponent elastic seismic wavefield using deep learning. *IEEE Geoscience and Remote Sensing Letters*, 20, 1–5. https://doi.org/10.1109/LGRS.2023.3250522

[23] Huang, X., & Alkhalifah, T. A. (2024). Microseismic source imaging using physics-informed neural networks with hard constraints. *IEEE Transactions on Geoscience and Remote Sensing*, 62, 1–11. https://doi.org/10.1109/TGRS.2024.3366449

[24] Okazaki, T., Ito, T., Hirahara, K., & Ueda, N. (2022). Physics-informed deep learning approach for modeling crustal deformation. *Nature Communications*, 13(1), 7092. https://doi.org/10.1038/s41467-022-34922-1

[25] Okazaki, T., Hirahara, K., & Ueda, N. (2024). Fault geometry invariance and dislocation potential in antiplane crustal deformation: physics-informed simultaneous solutions. *Progress in Earth and Planetary Science*, 11, 52. https://doi.org/10.1186/s40645-024-00654-7

[26] Okazaki, T., Hirahara, K., Ito, T., Kano, M. and Ueda, N., (2025), Physics-informed deep learning for forward and inverse modeling of inplane crustal deformation, *Journal of Geophysical Research: Machine Learning and Computation*, 2, e2024JH000474. https://doi.org/10.1029/2024JH000474

[27] Jagtap, A. D., & Karniadakis, G. E. (2020). Extended physics-informed neural networks (XPINNs): A generalized space-time domain decomposition based deep learning framework for nonlinear partial differential equations. *Communications in Computational Physics*, 28(5), 2002–2041. https://doi.org/10.4208/cicp.oa-2020-0164

[28] Shukla, K., Jagtap, A. D., & Karniadakis, G. E. (2021). Parallel physics-informed neural networks via domain decomposition. *Journal of Computational Physics*, 447, 110683. https://doi.org/10.1016/j.jcp.2021.110683

[29] Baydin, A. G., Pearlmutter, B. A., Radul, A. A., & Siskind J. M. (2018). Automatic differentiation in machine learning: a survey. *Journal of Machine Learning Research*, 18, 1–43 (2018).

[30] Krishnapriyan, A., Gholami, A., Zhe, S., Kirby, R., & Mahoney M. W. (2021). Characterizing possible failure modes in physics-informed neural networks. In *Advances in Neural Information Processing Systems*, 34, 26548–26560.

[31] Okada, Y., Internal deformation due to shear and tensile faults in a half-space. *Bulletin of the Seismological Society of America*, 82(2), 1018–1040 (1992). https://doi.org/10.1785/BSSA0820021018

[32] Aagaard, B. T., Knepley, M. G., & Williams, C. A. (2013). A domain decomposition approach to implementing fault slip in finite-element models of quasi-static and dynamic crustal deformation. *Journal of Geophysical Research: Solid Earth*, 118, 3059–3079. https://doi.org/10.1002/jgrb.50217

[33] Aoi, S., Kunugi, T., & Fujiwara, H. (2008). Trampoline effect in extreme ground motion. *Science*, 322(5902), 727–730.

[34] Ozawa, S., Imakiire, T., Tobita, M., Yarai, H., Nishimura, T., & Suito, H. (2008), Crustal deformation and seismic fault model of the Iwate-Miyagi nairiku earthquake in 2008, *GSI Journal*, 117, 79–80 (in Japanese).

[35] Ohta, Y., Ohzono, M., Miura, S., Iinuma, T., Tachibana, K., Takatsuka, K., Miyao, K., Sato, T., & Umino, N. (2008). Coseismic fault model of the 2008 Iwate-Miyagi Nairiku earthquake deduced by a dense GPS network. *Earth, planets and space*, 60, 1197–1201. https://doi.org/10.1186/BF03352878

[36] Suzuki, W., Aoi, S., & Sekiguchi, H. (2010). Rupture process of the 2008 Iwate–Miyagi Nairiku, Japan, earthquake derived from near-source strong-motion records. *Bulletin of the Seismological Society of America*, 100(1), 256–266. https://doi.org/10.1785/0120090043

[37] Koketsu, K., Miyake, H., & Suzuki, H. (2012). Japan integrated velocity structure model version 1. In *Proceedings of the 15th World Conference on Earthquake Engineering*, Lisbon.

[38] Kyriakopoulos, C., Masterlark, T., Stramondo, S., Chini, M., & Bignami, C. (2013). Coseismic slip distribution for the Mw 9 2011 Tohoku-Oki earthquake derived from 3-D FE modeling. *Journal of Geophysical Research: Solid Earth*, 118(7), 3837–3847. https://doi.org/10.1002/jgrb.50265

[39] Abe, T., Furuya, M., & Takada, Y. (2013). Nonplanar fault source modeling of the 2008 M w 6.9 Iwate–Miyagi Inland earthquake in Northeast Japan. *Bulletin of the Seismological Society of America*, 103(1), 507–518. https://doi.org/10.1785/0120120133

[40] Neyshabur, B., Tomioka, R., & Srebro, N. (2015). In search of the real inductive bias: On the role of implicit regularization in deep learning. In *International Conference on Learning Representations*.

[41] Karniadakis, G.E., Kevrekidis, I.G., Lu, L., Perdikaris, P., Wang, S., & Yang, L. (2021). Physics-informed machine learning. *Nature Reviews Physics*, 3, 422–440. https://doi.org/10.1038/s42254-021-00314-5

[42] Agata, R., Shiraishi, K. and Fujie, G., (2025), Physics-informed deep learning quantifies propagated uncertainty in seismic structure and hypocenter determination, *Scientific Reports*, 15, 1846. https://doi.org/10.1038/s41598-024-84995-9

[43] Agata, R., Shiraishi, K., & Fujie, G. (2023). Bayesian seismic tomography based on velocity-space Stein variational gradient descent for physics-informed neural network. *IEEE Transactions on Geoscience and Remote Sensing*, 61, 4506917, 1–17. https://doi.org/10.1109/TGRS.2023.3295414